\documentclass[prd,onecolumn,nofootinbib]{revtex4}

\usepackage{amsmath}
\usepackage{graphicx}
\usepackage{amsfonts}
\usepackage{latexsym}
\usepackage{bbold}
\usepackage{wasysym}
\usepackage{calligra}
\usepackage{float}
\usepackage{ulem}
\usepackage{inputenc}
\usepackage{xspace}
\usepackage{url}
\usepackage{epstopdf}
\usepackage{tikz}

\newcommand{\be}{\begin{equation}}
\newcommand{\ee}{\end{equation}}
\newcommand{\beq} {\begin{equation}}
\newcommand{\eeq} {\end{equation}}
\newcommand{\ba}{\begin{eqnarray}}
\newcommand{\ea}{\end{eqnarray}}

\usepackage{microtype}

\begin{document}

\title{The Cosmology of Quadratic Torsionful Gravity}
	
\author{Damianos Iosifidis$^\ast$, Lucrezia Ravera$^{\dag,\star}$}

\affiliation{$^\ast$Institute of Theoretical Physics, Department of Physics, Aristotle University of Thessaloniki, 54124 Thessaloniki, Greece. \\
			$^\dag$DISAT, Politecnico di Torino, Corso Duca degli Abruzzi 24, 10129 Torino, Italy. \\
			$^\star$INFN, Sezione di Torino, Via P. Giuria 1, 10125 Torino, Italy.}

\email{diosifid@auth.gr, lucrezia.ravera@polito.it}
	
\date{\today}
	
\begin{abstract}
		
We study the cosmology of a quadratic metric-compatible torsionful gravity theory in the presence of a perfect hyperfluid. The gravitational action is an extension of the Einstein-Cartan theory given by the usual Einstein-Hilbert contribution plus all the admitted quadratic parity even torsion scalars and the matter action also exhibits a dependence on the connection. The equations of motion are obtained by regarding the metric and the metric-compatible torsionful connection as independent variables. We then consider a Friedmann-Lema\^{i}tre-Robertson-Walker background, analyze the conservation laws, and derive the torsion modified Friedmann equations for our theory. Remarkably, we are able to provide exact analytic solutions for the torsionful cosmology.

\end{abstract}
	
\maketitle
	
\allowdisplaybreaks
	
%\newpage
	
\tableofcontents

%\newpage
	
\section{Introduction}\label{intro}

As it is well known, the development of Riemannian geometry led to the rigorous mathematical formulation of general relativity (GR). In spite of the great success and solid predictive power of GR in many contexts, it still falls short in explaining some of the current cosmological data. It does not properly explain the cosmological evolution at early times and is unable to predict a late time accelerated expansion. Consequently, diverse alternative modified theories of gravity have been proposed \cite{Clifton:2011jh}.
Among the various proposals, a particularly well motivated and promising setup in the spirit of gravity geometrization is that of non-Riemannian geometry \cite{eisenhart,schouten}, where the Riemannian assumptions of metric compatibility and torsionlessness of the connection are released and therefore non-vanishing torsion and nonmetricity are allowed along with curvature. Non-Riemannian effects, induced by the presence of torsion and non-metricity, are nowadays believed to have played a key role in particular in the very early Universe (see \cite{Puetzfeld:2004yg,Puetzfeld:2005af} and references therein).

Different restrictions of non-Riemannian geometry provide distinct frameworks for gravity theories formulations and the inclusion of torsion and non-metricity in gravitational theories has led to many fruitful applications in various areas of both mathematics and physics, among which, for instance, the ones recently presented in {\cite{Klemm:2018bil,Klemm:2020mfp,Klemm:2019izb,Klemm:2020gfm,Iosifidis:2020dck,Kranas:2018jdc,Pereira:2019yhu,Barrow:2019bvx,Guimaraes:2020drj,Iosifidis:2020zzp}}. In particular, in the cosmological context, in \cite{Iosifidis:2020zzp} the most general form of acceleration equation in the presence of torsion and non-metricity was derived and conditions under which torsion and non-metricity accelerate/decelerate the expansion rate of the Universe were discussed.
Let us also mention that imposing the vanishing of torsion and non-metricity one gets metric theories of which GR is a special case, whereas by demanding the vanishing of the curvature and non-metricity one is left with the standard teleparallel formulation \cite{aldrovandi}. Moreover, one could either set the curvature and torsion to zero while allowing for a non-vanishing non-metricity, which yields the symmetric teleparallel scheme \cite{Nester:1998mp,BeltranJimenez:2018vdo}, or fix just the curvature to zero getting a generalized teleparallel framework involving both torsion and non-metricity \cite{Jimenez:2019ghw}. 
On the other hand, one may also impose no constraint on such geometric objects. This is the non-Riemannian scenario where Metric-Affine Gravity (MAG) theories are developed. The literature on the subject is huge. For an exhaustive review of the geometrical theoretical background on MAG we refer the reader to e.g. \cite{Hehl:1994ue,Hehl:1999sb,Iosifidis:2019jgi}.
In the metric-affine approach the metric and the connection are considered as independent fields and the matter Lagrangian depends on the connection as well. In this framework, the theory is assumed to have, in principle, a non-vanishing hypermomentum tensor \cite{Hehl:1976kt} encompassing the microscopic characteristics of matter such as spin, dilation, and shear \cite{Hehl:1994ue}.

What is more, in the framework of non-Riemannian geometry, where the presence of extra degrees of freedom with respect to GR is due to torsion and non-metricity of spacetime which are linked to the microstructure of matter, fluid carrying hypermomentum turns out to be very appealing. In particular, diverse hyperfluid models have proved to have relevant applications especially in cosmology, such as the ones given in \cite{Weyssenhoff:1947iua,Obukhov:1993pt,Obukhov:1996mg,Babourova:1995fv,Babourova:2004xx,Ray:1982qr,Smalley,Iosifidis:2020gth,Iosifidis:2020upr,Iosifidis:2021nra}. In particular, in \cite{Iosifidis:2021nra} the perfect (ideal) hyperfluid model representing the natural generalization of the classical GR perfect fluid structure has been formulated and analyzed. 

Motivated by the prominent and intriguing role of non-Riemannian geometry and hyperfluids in the cosmological scenario, in the present paper we study the cosmology of a quadratic torsionful gravity theory given by the Einstein-Hilbert (EH) contribution plus all the admitted quadratic parity even torsion scalars (see also \cite{Baekler:2011jt}) and in the presence of a perfect hyperfluid. We restrict ourselves to the case of vanishing non-metricity while allowing for a non-vanishing torsion and let the matter action also exhibit a dependence on the connection. 

The remaining of this paper is structured as follows: In Section \ref{geomsetup} we briefly review the geometric setup and in Section \ref{hmemrev} we give a short account of energy-momentum and hypermomentum tensors. Subsequently, in Section \ref{thetheory} we write our quadratic torsionful gravity theory and derive its field equations. We work in a first order formalism, where the metric and the affine connection are treated as independent variables. The theory and the aforementioned general analysis is developed in $n$ spacetime dimensions, whereas we restrict ourselves to the case $n=4$ when studying solutions. Section \ref{cosmqtt} is devoted to the study of the cosmology of the theory. Here we first discuss the torsion degrees of freedom in a Friedmann-Lema\^{i}tre-Robertson-Walker (FLRW) spacetime and recall the notion of perfect hyperfluid together with its properties. Then we analyze the field equations, conservation laws, and torsion modified Friedmann equations for our torsionful model. Finally, in Section \ref{solsec} we provide exact analytic solutions for such torsionful cosmology. In Section \ref{conclusions} we discuss our results and possible future developments. Useful formulas and conventions are collected in Appendix \ref{appa}.

\section{Review of the geometric setup}\label{geomsetup}

Let us start with a brief review of the geometric setup. We will adopt the same notation and conventions of Ref. \cite{Iosifidis:2019jgi}, to which we refer the reader for more details. 
We consider the framework of non-Riemannian geometry, endowed with a metric $g_{\mu \nu}$ and an independent affine connection ${\Gamma^\lambda}_{\mu \nu}$. 
The generic decomposition of an affine connection reads
\beq\label{gendecompaffconn}
{\Gamma^\lambda}_{\mu\nu} = \tilde{\Gamma}^\lambda_{\phantom{\lambda}\mu\nu} + {N^\lambda}_{\mu\nu}\,,
\eeq
where the distortion tensor ${N^\lambda}_{\mu\nu}$ (non-Riemannian contribution to the affine connection) and the Levi-Civita connection $\tilde{\Gamma}^\lambda_{\phantom{\lambda}\mu\nu}$ (Riemannian contribution) are respectively given by
\beq\label{distortion}
{N^\lambda}_{\mu\nu} = \underbrace{\frac12 g^{\rho\lambda}\left(Q_{\mu\nu\rho} + Q_{\nu\rho\mu}
- Q_{\rho\mu\nu}\right)}_{\text{deflection {(or disformation)}}} - \underbrace{g^{\rho\lambda}\left(S_{\rho\mu\nu} +
S_{\rho\nu\mu} - S_{\mu\nu\rho}\right)}_{\text{contorsion} \, := \, {K^\lambda}_{\mu \nu}}\,,
\eeq
\beq\label{lcconn}
\tilde{\Gamma}^\lambda_{\phantom{\lambda}\mu\nu} = \frac12 g^{\rho\lambda}\left(\partial_\mu 
g_{\nu\rho} + \partial_\nu g_{\rho\mu} - \partial_\rho g_{\mu\nu}\right)\,.
\eeq
In eq. \eqref{distortion}, ${S_{\mu\nu}}^\rho$ is the Cartan torsion tensor,
\beq
{S_{\mu\nu}}^\lambda := {\Gamma^\lambda}_{[\mu\nu]}\,, \label{torsdef}
\eeq
whose trace is given by
\beq\label{tortracedef}
S_\mu := {S_{\mu \lambda}}^\lambda \,.
\eeq
On the other hand, $Q_{\lambda\mu\nu}$ is the nonmetricity tensor, defined as
\beq
Q_{\lambda\mu\nu}:= -\nabla_\lambda g_{\mu\nu} = 
-\partial_\lambda g_{\mu\nu} + {\Gamma^\rho}_{\mu\lambda} g_{\rho\nu} +
{\Gamma^\rho}_{\nu\lambda}g_{\mu\rho} \,.
\eeq
In the sequel we will focus on the case of a metric-compatible torsionful affine connection, namely we will consider vanishing non-metricity and non-vanishing torsion.
Our definition for the covariant derivative $\nabla$, associated with a metric-compatible torsionful affine connection $\Gamma$, acting on a vector is
\beq
\nabla_\mu u^\lambda = \partial_\mu u^\lambda + {\Gamma^\lambda}_{\nu \mu} u^\nu \,.
\eeq
The curvature tensor is defined by
\beq\label{curvtensdef}
{R^\mu}_{\nu \alpha \beta} := 2 \partial_{[\alpha} {\Gamma^\mu}_{|\nu|\beta]} + 2 {\Gamma^\mu}_{\rho[\alpha} {\Gamma^\rho}_{|\nu|\beta]} 
\eeq
and we also have the following contractions:
\begin{align}
& R_{\nu \beta} := {R^\mu}_{\nu \mu \beta} \,, \label{Riccitens} \\
& \hat{R}_{\alpha \beta} := {R^\mu}_{\mu \alpha \beta} = 0 \,, \label{homot} \\
& \check{R}^\lambda_{\phantom{\lambda} \alpha} := {R^\lambda}_{\mu \nu \alpha} g^{\mu \nu} \label{coRiccitens} \,. 
\end{align}
The tensor in \eqref{Riccitens} is the Ricci tensor of $\Gamma$, while in \eqref{homot} we have the so-called homothetic curvature which vanishes for metric-compatible affine connections, and in \eqref{coRiccitens} we have introduced a third tensor that is sometimes referred to as the co-Ricci tensor in the literature. In particular, for metric-compatible affine connections we have $\check{R}_{\mu \nu}=-R_{\mu \nu}$ (see also \cite{Iosifidis:2019jgi} for details). A further contraction gives us the Ricci scalar of $\Gamma$, which is uniquely defined, since
\beq 
R := R_{\mu \nu} g^{\mu \nu} = - \check{R}_{\mu \nu} g^{\mu \nu} \,.
\eeq
Let us also mention that plugging the decomposition \eqref{gendecompaffconn} into the definition of the curvature tensor \eqref{curvtensdef} one can prove that
\beq
{R^\mu}_{\nu \alpha \beta} = \tilde{R}^\mu_{\phantom{\mu} \nu \alpha \beta} + 2 \tilde{\nabla}_{[\alpha} {N^\mu}_{|\nu|\beta]} + 2 {N^\mu}_{\lambda|\alpha} {N^\lambda}_{|\nu|\beta]} \,,
\eeq
where $\tilde{\nabla}$ denotes the Levi-Civita covariant derivative. Moreover, the torsion can be derived from the distortion tensor through the relation
\beq
S_{\mu \nu \alpha} = N_{\alpha [\mu \nu]} \,.
\eeq
The variation of the torsion with respect to the metric and the connection (see e.g. \cite{Iosifidis:2019jgi}) reads, respectively,
\beq
\delta_g {S_{\mu\nu}}^\alpha = 0\,, \label{deltagt}
\eeq
\beq
\delta_\Gamma {S_{\alpha\beta}}^\lambda = \delta^{[\mu}_{\alpha}\delta^{\nu]}_{\beta}
\delta {\Gamma^\lambda}_{\mu\nu}\,. \label{deltaconnt}
\eeq
These formulas are particularly useful to reproduce the calculations in the sequel.

\section{Hypermomentum and energy-momentum tensors}\label{hmemrev}

In this section we give a short account of energy-momentum and hypermomentum tensors, following the same lines of \cite{Iosifidis:2020gth}. Here we shall restrict ourselves to the metric-compatible torsionful case.

In our setup we consider the action to be a functional of the metric, the independent metric-compatible torsionful connection, and the matter fields, that is to say
\beq \label{magact}
S[g,\Gamma,\varphi] = S_{\text{G}}[g,\Gamma] + S_{\text{M}}[g,\Gamma,\varphi] \,,
\eeq
where 
\beq
S_{\text{G}}[g,\Gamma] = \frac{1}{2\kappa} \int d^n x \sqrt{-g} \mathcal{L}_{\text{G}} (g,\Gamma) 
\eeq
and 
\beq
S_{\text{M}}[g,\Gamma,\varphi] = \int d^n x \sqrt{-g} \mathcal{L}_{\text{M}} (g,\Gamma,\varphi) 
\eeq
represent, respectively, the gravitational sector and the matter one. In the former $\kappa=8\pi G$ is the gravitational constant, while in the latter $\varphi$ collectively denotes the matter fields. Let us mention that the action \eqref{magact} also depends on the derivatives of the metric and connection. Here we are suppressing the aforesaid dependence for simplicity.

Then, we define as usual the metrical (symmetric) energy-momentum tensor (MEMT)
\beq
T_{\mu \nu} := - \frac{2}{\sqrt{-g}} \frac{\delta S_{\text{M}}}{\delta g^{\mu \nu}} = - \frac{2}{\sqrt{-g}} \frac{\delta (\sqrt{-g} \mathcal{L}_{\text{M}})}{\delta g^{\mu \nu}}
\eeq
and the hypermomentum tensor (HMT) \cite{Hehl:1976kt}
\beq
{\Delta_\lambda}^{\mu \nu} := - \frac{2}{\sqrt{-g}} \frac{\delta S_{\text{M}}}{\delta {\Gamma^\lambda}_{\mu \nu}} = - \frac{2}{\sqrt{-g}} \frac{\delta (\sqrt{-g} \mathcal{L}_{\text{M}})}{\delta {\Gamma^\lambda}_{\mu \nu}} \,,
\eeq
which encompasses matter microstructure \cite{Hehl:1994ue}. Now, note that if one works in the equivalent formalism based on the vielbeins ${e_{\mu}}^c$ and spin connection $\omega_{\mu a b}$, then the so-called canonical energy-momentum tensor (CEMT) is defined by
\beq\label{cemt}
{t^\mu}_c := \frac{1}{\sqrt{-g}} \frac{\delta S_{\text{M}}}{\delta {e_\mu}^c} \,,
\eeq
which, in general, is not symmetric. Here we use Latin letters to denote Lorentz indices, that is tangent indices. The usual relation $g_{\mu \nu}= {e_\mu}^a {e_\nu}^b \eta_{ab}$ connecting metric and vielbeins holds, where $\eta_{ab}$ is the tangent space flat Minkowski metric. Our conventions are given in Appendix \ref{appa}. 
The CEMT is not independent of the MEMT and HMT (see also \cite{Hehl:1994ue}). Indeed, one can prove that the following relation holds:
\beq\label{cemt1}
{t^\mu}_\lambda := \frac{1}{\sqrt{-g}} \frac{\delta S_{\text{M}}}{\delta {e_\lambda}^c} {e_\lambda}^c = {T^\mu}_\lambda - \frac{1}{2 \sqrt{-g}} \hat{\nabla}_\nu \left( \sqrt{-g} {\Delta_\lambda}^{\mu \nu} \right) \,,
\eeq
where we have also exploited the identity
\beq
\nabla_\nu {e_\mu}^a = 0 = \partial_\nu {e_\mu}^a - {\Gamma^\rho}_{\mu \nu} {e_\rho}^a + \omega^{\phantom{\nu} a}_{\nu \phantom{a} b} {e_\mu}^b  \,,
\eeq
connecting the two formalisms, and we have defined
\beq\label{hatnabla}
\hat{\nabla}_\nu := 2 S_\nu - \nabla_\nu \,.
\eeq
Observe that for matter with no microstructure ($\Delta_{\alpha \mu \nu}\equiv 0$) the CEMT and MEMT coincide. Furthermore, note that from eq. \eqref{cemt1} one can obtain the conservation law for spin \cite{Obukhov:2014nja}, which, in the metric-compatible torsionful case, reads
\beq
2 t_{[\mu \nu]} = \frac{1}{\sqrt{-g}} \hat{\nabla}_\alpha \left( \sqrt{-g} {\tau_{\mu \nu}}^\alpha \right) \,,
\eeq
where
\beq
{\tau_{\mu \nu}}^\alpha := {\Delta_{[\mu \nu]}}^\alpha \,.
\eeq
Additionally, upon contraction of $\mu,\lambda$ in \eqref{cemt1}, one gets the trace relation
\beq\label{tracerelation}
t = T + \frac{1}{2 \sqrt{-g}} \partial_\nu \left( \sqrt{-g} \Delta^\nu \right) \,,
\eeq
with
\beq
t := {t^\mu}_\mu \,, \quad T := {T^\mu}_\mu \,, \quad \Delta^\nu := {\Delta_\lambda}^{\lambda \nu} \,.
\eeq
From eq. \eqref{tracerelation} one can notice that for specific matter types the following relations hold true:
\begin{align}
T = 0 \quad & \leftrightarrow \quad 2t = \frac{1}{\sqrt{-g}} \partial_\nu \left( \sqrt{-g} \Delta^\nu \right) \,, \\
t = 0 \quad & \leftrightarrow \quad 2T = - \frac{1}{\sqrt{-g}} \partial_\nu \left( \sqrt{-g} \Delta^\nu \right) \,, \\
t = T \quad & \leftrightarrow \quad \partial_\nu \left( \sqrt{-g} \Delta^\nu \right)  = 0 \,,
\end{align}
corresponding, respectively, to the case of conformally invariant, frame rescalings invariant, and special projective transformations invariant theories (see \cite{Iosifidis:2018zwo} for details on such models).

\section{The theory}\label{thetheory}
	
We consider an extension of the Einstein-Cartan theory, including also the three torsion (parity even) quadratic terms that are allowed by dimensional analysis.\footnote{These terms have exactly the same dimension as $R$, that is $[L^{-2}]$, and therefore their inclusion is well motivated.} As we shall show, their presence is rather essential in order to obtain non-trivial dynamics for the torsion variables.\footnote{In \cite{Kranas:2018jdc} {(see also \cite{Pereira:2019yhu,Barrow:2019bvx,Guimaraes:2020drj}, where the form of torsion presented in \cite{Kranas:2018jdc} was constrained)}, where only the Ricci scalar was present in the gravitational action, the conservation law for hypermomentum was trivialized leaving, therefore, the torsion function fully undetermined. It will be shown in the sequel that the very presence of the quadratic torsion terms fixes this indeterminacy, providing exact evolution laws for all variables.} Then, our extended quadratic torsionful action involves three parameters and reads
\beq \label{genact}
S[g,\Gamma,\varphi] = \frac{1}{2 \kappa}\int d^{n}x \sqrt{-g} \Big[ R+ b_{1}S_{\alpha\mu\nu}S^{\alpha\mu\nu} +
b_{2}S_{\alpha\mu\nu}S^{\mu\nu\alpha} + 	b_{3}S_{\mu}S^{\mu}  \Big] +S_{\text{hyp}} \,,
\eeq
where $b_{1}$, $b_{2}$, and $b_{3}$ are dimensionless parameters. $S_{\text{hyp}}$ denotes the matter part which we assume to be that of a perfect hyperfluid. Note that the above action is a special case of the more general gravitational theory involving both torsion and non-metricity quadratic parity even and parity odd terms \cite{Iosifidis:2019jgi}. Moreover, the action \eqref{genact} has been considered also in \cite{Baekler:2011jt} in a different context and in the presence of a cosmological constant. In this regard, let us mention here that the inclusion of a cosmological constant term in \eqref{genact} would just imply a further contribution to the metric field equations we are going to analyze, while the connection field equations would not be modified.
	
Let us now derive the field equations of the theory \eqref{genact}. Variation with respect to the metric gives
\beq 
R_{(\mu\nu)}-\frac{R}{2}g_{\mu\nu}-\frac{\mathcal{L}_{2}}{2}g_{\mu\nu} + B_{\mu\nu}=\kappa T_{\mu\nu} \,, \label{metricf}
\eeq
where we have defined	
\beq
B_{\mu\nu}=B_{\nu \mu}:=b_{1}(2S_{\nu\alpha\beta}{S_{\mu}}^{\alpha\beta}-S_{\alpha\beta\mu}{S^{\alpha\beta}}_{\nu})-b_{2}S_{\nu\alpha\beta}{S_{\mu}}^{\beta\alpha}+b_{3}S_{\mu}S_{\nu} 
\eeq
and
\beq
\mathcal{L}_{2}:=	b_{1}S_{\alpha\mu\nu}S^{\alpha\mu\nu} +
b_{2}S_{\alpha\mu\nu}S^{\mu\nu\alpha} + b_{3}S_{\mu}S^{\mu} \,.
\eeq
In addition, varying the action with respect to the metric-compatible but torsionful connection ${\Gamma^\lambda}_{\mu \nu}$ we get the field equations
\beq
{P_{\lambda}}^{\mu\nu}+{\Psi_{\lambda}}^{\mu\nu}=\kappa {\Delta_{\lambda}}^{\mu\nu} \,, \label{conn2}
\eeq
where
\beq
{P_{\lambda}}^{\mu\nu}=2(S_{\lambda}g^{\mu\nu}-S^{\mu}\delta_{\lambda}^{\nu}+g^{\mu\sigma}{S_{\sigma\lambda}}^{\nu})
\eeq
is the metric-compatible torsionful Palatini tensor, which fulfills ${P_{\mu}}^{\mu\nu}\equiv0$, and where we have defined
\beq
{\Psi_{\lambda}}^{\mu\nu}:= 2 b_{1}{S^{\mu\nu}}_{\lambda}+2 b_{2}{S_{\lambda}}^{[\mu\nu]}+2b_{3}S^{[\mu}\delta^{\nu ]}_{\lambda} \,.
\eeq
In what follows we will analyze the cosmology of this quadratic torsionful gravity theory.
	
\section{Cosmology with quadratic torsion terms}\label{cosmqtt}

In this section we move on to the study of the cosmology of the theory \eqref{genact}. 
To pursue this aim, we shall consider a flat FLRW spacetime with the usual Robertson-Walker line element
\beq
ds^{2}=-dt^{2}+a^{2}\delta_{ij}dx^{i}dx^{j} \,,
\eeq
where $a(t)$ is the cosmic scale factor and $i,j=1,2,\ldots,n-1$. In addition we let $u^\mu$ represent the normalized $n$-velocity field of a given fluid which in co-moving coordinates is expressed as $u^\mu= \delta^\mu_0=(1,0,0,\ldots,0)$, $u_\mu u^\mu=-1$. Accordingly, we define in the usual way the projector tensor
\beq\label{projop}
h_{\mu \nu}:= g_{\mu \nu} + u_\mu u_\nu \,,
\eeq
which project objects on the space orthogonal to $u^\mu$. We also define the temporal derivative
\beq\label{tempdev}
\dot{}=u^\alpha \nabla_\alpha \,.
\eeq
The projection operator \eqref{projop} and the temporal derivative \eqref{tempdev} constitute together a $1+(n-1)$ spacetime split.

\subsection{Perfect cosmological hyperfluid}\label{hyprev}
	
As we have already mentioned in the introduction, a hyperfluid is a classical continuous medium carrying hypermomentum. The general formulation of perfect hyperfluid generalizing the classical perfect fluid notion of GR has been recently presented in \cite{Iosifidis:2021nra} by first giving its physical definition and
later using the appropriate mathematical formulation in order to extract its energy tensors by demanding spatial isotropy. In our study we consider such perfect hyperfluid model in a homogeneous cosmological setting.\footnote{That is, we also demand homogeneity along with isotropy.}

As shown in \cite{Iosifidis:2021nra}, the description of the perfect hyperfluid is given by the energy related tensors
\beq\label{metrenmomform}
T_{\mu \nu} = \rho u_\mu u_\nu + p h_{\mu \nu} \,,
\eeq
\beq\label{canonenmomform}
t_{\mu \nu} = \rho_c u_\mu u_\nu + p_c h_{\mu \nu} \,,
\eeq
\beq\label{hypermomform}
\Delta^{(n)}_{\alpha \mu \nu} = \phi(t) h_{\mu \alpha} u_\nu + \chi(t) h_{\nu \alpha} u_{\mu} + \psi(t) u_{\alpha} h_{\mu \nu} + \omega(t) u_\alpha u_\mu u_\nu + \delta^n_4 \varepsilon_{\alpha \mu \nu \rho} u^\rho \zeta(t) \,,
\eeq
all of them respecting spatial isotropy and subject to certain conservation laws (see discussion below). In the hyperfluid MEMT \eqref{metrenmomform}, $\rho$ and $p$ are, as usual, the density and pressure of the perfect fluid component of the hyperfluid, while, in the hyperfluid CEMT \eqref{canonenmomform}, $\rho_c$ and $p_c$ are, respectively, the canonical density and canonical pressure of the hyperfluid. On the other hand, the variables $\phi$, $\chi$, $\psi$, $\omega$, and $\zeta$ in the hypermomentum \eqref{hypermomform} characterize the microscopic properties of the fluid which, upon using the connection field equations, act as the sources of the torsionful non-Riemannian background. The aforementioned conservation laws for the perfect hyperfluid in the case in which the non-metricity is set to zero while the torsion is non-vanishing read as follows:
\beq\label{conslawshyp1}
\frac{1}{\sqrt{-g}} \hat{\nabla}_\mu \left( \sqrt{-g} {t^\mu}_\alpha \right) = \frac{1}{2} \Delta^{\lambda \mu \nu} R_{\lambda \mu \nu \alpha} + 2 S_{\alpha \mu \nu} t^{\mu \nu} \quad \rightarrow \quad \tilde{\nabla}_\mu {t^\mu}_\alpha = \frac{1}{2} \Delta^{\lambda \mu \nu} R_{\lambda \mu \nu \alpha} \,,
\eeq
\beq\label{conslawshyp2}
{t^\mu}_\lambda = {T^\mu}_\lambda - \frac{1}{2 \sqrt{-g}} \hat{\nabla}_\nu \left( \sqrt{-g} {\Delta_\lambda}^{\mu \nu} \right) \,,
\eeq
Recall that $\tilde{\nabla}$ denotes the Levi-Civita covariant derivative. Notice that \eqref{conslawshyp2} is exactly the same relation \eqref{cemt1} we have previously obtained connecting the three energy related tensors.\footnote{Working in the language of differential forms the second conservation law comes from the $\mathrm{GL}(n,\mathbb{R})$ invariance of the matter part. On the other hand, the first one is obtained from diffeomorphism invariance. We refer the interested reader to \cite{Iosifidis:2021nra} for details.} In addition, as we can see from \eqref{conslawshyp1}, the canonical energy-momentum tensor naturally couples to torsion. Eqs. \eqref{conslawshyp1} and \eqref{conslawshyp2} will be fundamental in the study of the cosmology of our theory.

\subsection{Cosmology with torsion}

Now we need the most general form of torsion that can be written in a homogeneous and isotropic space. In such a space the torsion has at most two degrees of freedom in $n=4$ and a single one for $n \neq 4$ \cite{mtsamparlis}, and it can be written in an explicitly covariant fashion as (see also \cite{Iosifidis:2020gth,Iosifidis:2020zzp})
\beq
S_{\mu\nu\alpha}^{(n)}=2u_{[\mu}h_{\nu]\alpha}\Phi(t)+\varepsilon_{\mu\nu\alpha\rho}u^{\rho}P(t)\delta^n_4 \,, \label{torsionform}
\eeq
where $\varepsilon_{\mu\nu\alpha\rho}$ is the Levi-Civita tensor and $\delta^{n=4}_4=1$, otherwise it is zero. Here the upper label $(n)$ is used to denote that we are considering $n$ spacetime dimensions. Eq. \eqref{torsionform} also implies
\beq
S_\alpha = (n-1) \Phi u _\alpha \,,
\eeq
and the following relations hold:
\begin{align}
& S_{\mu \nu \alpha} S^{\mu \nu \alpha} = - 2 (n-1) \Phi^2 + 6 P^2 \delta^n_4 \,, \\
& S_{\mu \nu \alpha} S^{\alpha \mu \nu} = (n-1) \Phi^2 + 6 P^2 \delta^n_4 \,, \\
& S_\mu S^\mu = - (n-1)^2 \Phi^2 \,,
\end{align}
which imply, in particular,
\beq\label{torsrel}
(n-1) S_{\mu \nu \alpha} S^{\mu \nu \alpha} - (n-1) S_{\mu \nu \alpha} S^{\alpha \mu \nu} - 3 S_{\mu} S^{\mu} = 0 \,,
\eeq
indicating that actually only two out of the three torsion scalars are independent.

Consequently, the distortion tensor takes the form
\beq
N_{\alpha\mu\nu}^{(n)}=-\left(S_{\alpha \mu \nu}^{(n)} + S_{\alpha \nu \mu}^{(n)}-S_{\mu \nu \alpha}^{(n)}\right)=X(t)u_{\alpha}h_{\mu\nu}+Y(t)u_{\mu}h_{\alpha\nu} +\varepsilon_{\alpha\mu\nu\lambda}u^{\lambda}W(t)\delta^n_4 \,.
\eeq
Note that the functions determining the distortion are linearly related with the functions of torsion. This can be shown by using the fact that
\beq
S_{\mu\nu\alpha}^{(n)}=N_{\alpha[\mu\nu]}^{(n)} \,,
\eeq	
which results in the relations
\beq
2(X+Y)=0 \,, \quad 2\Phi = Y \,, \quad P = W \,,
\eeq
or, inverting them,
\beq
W=P \,, \quad Y=2\Phi \,, \quad X=-2\Phi \,.
\eeq
Therefore, non-Riemannian effects driven by torsion can be parametrized using either the set $\lbrace \Phi, P \rbrace$ or the set $\lbrace X,Y,W \rbrace \rightarrow \lbrace Y,W \rbrace$ (in fact, notice that we have $Y=-X=2\Phi$). Both of these sets will be related to the set of hypermomentum sources by means of the connection field equations of the theory. Nevertheless, in what follows we shall use the former, which provides a more transparent geometrical meaning with respect to the latter.

Let us also recall that the hyperfluid energy related tensors take the form \eqref{metrenmomform}, \eqref{canonenmomform}, and \eqref{hypermomform}. Moreover, since we are considering a metric-compatible setup (that is to say vanishing non-metricity), we will also have $\omega=0$, since $\omega u_\alpha u_\mu u_\nu$, being totally symmetric, can only excite non-metric degrees of freedom, which are absent here.

\subsection{Analysis of the connection field equations}

Using the information collected above and contracting the connection field equations \eqref{conn2} independently in $\mu,\lambda$, then in $\nu,\lambda$, and finally with $g_{\mu\nu}$, we get the following three equations:
\beq
-\Big[ 2b_{1} -b_{2} +(n-1)b_{3} \Big] (n-1) \Phi=\kappa \Big[ (n-1)\phi-\omega \Big] \,,
\eeq
\beq
\Big[ -2(n-2)+2 b_{1}- b_{2}+(n-1)b_{3} \Big](n-1)\Phi=\kappa\Big[ (n-1)\chi -\omega \Big] \,,
\eeq
\beq
2(n-2)(n-1)\Phi=\kappa \Big[ (n-1)\psi -\omega \Big]	 \,.
\eeq
Moreover, the contraction of the connection field equations with $u^{\lambda}u_{\mu}u_{\nu}$ gives the constraint $\omega=0$, which we already anticipated, since this part of hypermomentum, being totally symmetric, can only excite the non-metric degrees of freedom which are absent here. In addition, the pseudo-scalar torsion mode is obtained by taking the totally antisymmetric part of \eqref{conn2}, which yields
\beq\label{Pzeta}
2(b_1 +b_2 -1)P =\kappa \zeta \,.
\eeq
Let us observe that the assumption $b_1 + b_2 \neq 1$ is crucial here, since otherwise one would face the constraint $\zeta(t)=0$ on the sources which would then make $P(t)$ arbitrary, signaling a problematic (unphysical) theory. It is therefore natural to assume that $b_{1} + b_2 \neq 1$. Combining the above equations we have
\begin{align}
& \kappa \phi =-\Big[ 2b_{1} -b_{2} +(n-1)b_{3} \Big]\Phi \,, \nonumber \\
& \kappa \psi = 2(n-2)\Phi \,, \nonumber \\
& \kappa\chi=-2(n-2)\Phi+\Big[ 2b_{1}-b_{2}+(n-1)b_{3} \Big]\Phi=-\psi-\phi \,, \\ 
& \omega=0 \,, \nonumber \\
& 2(b_1 + b_2 -1)P =\kappa \zeta \,. \nonumber
\end{align}
Note that the latter imply that the hypermomentum variables are related to each other, since it is evident that the following relations hold true:
\beq\label{chi}
\chi=\left[ \frac{2(n-2)}{2 b_{1} -b_{2} +(n-1)b_{3}}-1\right] \phi \,,
\eeq
\beq\label{psi}
\psi=- \frac{2(n-2)}{2 b_{1} -b_{2}+(n-1)b_{3}}\phi \,, 
\eeq
with
\beq
\chi+\psi=-\phi \,.
\eeq
The dynamics is therefore contained in $\phi$. In the above the assumption $2 b_{1}-b_{2}+(n-1)b_{3}\neq 0$ has been made. The latter is crucial in order to obtain non-trivial solutions. Moreover, note that if the quadratic terms are switched off, that is if $b_{1}=0=b_{2}=b_{3}$, it follows that both $\phi=0$ and $\omega=0$, as seen from the above. Then, in such a case, there are no evolution equations for the hypermomentum variables and subsequently $\Phi$ remains undetermined (this was in fact the case in \cite{Kranas:2018jdc}). 
		
\subsection{Conservation laws}
Using \eqref{metrenmomform}, \eqref{canonenmomform}, and \eqref{hypermomform}, one can easily prove that the continuity equation from \eqref{conslawshyp1} in the present case reads
\beq
\dot{\rho}_{c}+(n-1)H(\rho_{c}+p_{c})=\frac{1}{2}(\psi-\chi)R_{\mu\nu}u^{\mu}u^{\nu} \,,	
\eeq	
where, as usual, $H:= \frac{\dot{a}}{a}$ is the Hubble parameter. On the other hand, taking the $00$ and $ij$ components of the conservation law \eqref{conslawshyp2} we obtain the evolution equations for the hypermomentum variables, which result to be given by
\beq
\dot{\omega}+(n-1)H(\chi+\psi+\omega)+(n-1)( \psi X-\chi Y)=2(\rho_{c}-\rho) \,, \label{hyper2}
\eeq
\beq
\dot{\phi}+(n-1)H \phi +H(\chi +\psi) +\psi X- \chi Y=2 (p_{c}-p) \,. \label{hyper1}
\eeq
Moreover, since in our case $\omega=0$, the first becomes
\beq\label{hyper1new}
(n-1)(\psi +\chi)(H-Y)=2 (\rho_{c}-\rho) \,,
\eeq	
where we have also used the fact that $Y=-X$. 

Let us now see what happens if we assume that our hyperfluid is
of the hypermomentum preserving type \cite{Iosifidis:2020gth}. In this case the metrical and canonical energy momentum tensors coincide
and as a result $\rho_{c}=\rho$ as well as $p_{c}=p$. Then, the above equation becomes
\beq
(n-1) (\psi +\chi)(H-Y)=0 \label{HY}
\eeq
and therefore it follows that either
\beq
\psi+\chi=0	\label{s1}
\eeq
or
\beq
H-Y=0 \,. \label{incompr}
\eeq
Remarkably, each of the above constraints has a direct physical interpretation. Indeed, as it can be seen from the hypermomentum decomposition \eqref{hypermomform}, the combination $\psi+\chi$ appears in the shear part of hypermomentum. Therefore, eq. \eqref{s1} is related to the vanishing of one of the shear sources. On the other hand, one can trivially verify that
\beq
\nabla_{i}u^{i}=(n-1)(H-Y) \,.
\eeq
Hence, eq. \eqref{incompr} turns out to imply that the hyperfluid is incompressible, $\nabla_{i}u^{i}=0$. Thus, we see that eq. \eqref{HY} has a very clear interpretation, meaning that the fluid must either have one shear part vanishing or it should be incompressible. 

Let us now go back to the analysis of eq. \eqref{hyper1}. Using \eqref{HY} and continuing to consider the case of hypermomentum preserving hyperfluid, eq. \eqref{hyper1} simplifies to
\beq
\dot{\phi}+(n-1)H \phi=0	\,.
\eeq
Had we not assumed hypermomentum preserving configuration, this would generalize to
\beq
\dot{\phi}+(n-1)H \phi=\frac{2}{(n-1)}\Big[ -(\rho_{c}-\rho)+(n-1)(p_{c}-p) \Big] \,.
\eeq
In order to keep the following discussion as general as possible, we shall not assume, at this point, that the hyperfluid is {of hypermomentum preserving type. We will further come back to this special case with some observations at the end of Section \ref{solsec}, where we will study solutions of our model.}

\subsection{Torsion modified Friedmann equations}

We are now in a position to derive the torsionful Friedmann equations. Taking the $00$ components of the metric field equations \eqref{metricf}, after some calculations (see Appendix \ref{appa} for a collection of useful formulas we have derived and exploited in our computations), we finally find
\beq
H^{2}=-\frac{1}{(n-2)}\Big[ 2b_{1}-b_{2}+(n-1)b_{3}+4(n-2) \Big]\Phi^{2}+4 H\Phi +(1-b_{1}-b_{2})P^{2}\delta^n_4 + \frac{2 \kappa}{(n-1)(n-2)}\rho \,. \label{Friedmann10}
\eeq
Note that when the quadratic torsion terms are absent the above reduces to
\beq
H^{2}=-4 \Phi^{2}+4 H\Phi+P^{2}\delta^n_4 +\frac{2 \kappa}{(n-1)(n-2)}\rho \,,
\eeq
which is in perfect agreement with \cite{Kranas:2018jdc,Iosifidis:2020gth}, as expected. The second Friedmann equation (also known as acceleration equation) can be obtained by combining the $00$ and the $ij$ components of the metric field equations. This would require some cumbersome calculations, but eventually there exists a much simpler road. Indeed, in \cite{Iosifidis:2020zzp} the most general form of the acceleration equation was derived and, in the case of vanishing non-metricity we are considering, it takes the form
\beq
\frac{\ddot{a}}{a}=-\frac{1}{(n-1)}R_{\mu\nu}u^{\mu}u^{\nu}+2 \left( \frac{\dot{a}}{a} \right) \Phi +2\dot{\Phi} \,. \label{acc}
\eeq
Therefore, we can find the second Friedmann equation by just computing the piece $R_{\mu\nu}u^{\mu}u^{\nu}$ from the metric field equations in the present case (see Appendix \ref{appa} for details). We get
\beq
\frac{\ddot{a}}{a}=-\frac{\kappa}{(n-1)(n-2)}\Big[ (n-3)\rho+(n-1)p \Big]+\Big[ 2b_{1}-b_{2}+(n-1)b_{3}\Big]\Phi^{2}+2 H \Phi+2 \dot{\Phi} \,. \label{acc20}
\eeq
Observe that in the case $n=3$ the $\rho$ contribution disappear. On the other hand, in $n=4$ all terms survive, and this is the case we will restrict to in the following, where we are going to discuss solutions of our cosmological theory.	
	
\section{Solutions}\label{solsec}

In this section we derive exact analytic solutions of our torsionful cosmological model. Before proceeding in this direction, let us just recall that, as we have shown, since \eqref{torsrel} holds true not all the three quadratic torsion invariants are linearly independent. This means that we may set one of the $b_a$ ($a=1,2,3$) to zero, which would amount to a renaming of the $b_a$ as they appear in \eqref{genact}. We choose to set $b_{2}=0$. Then, our cosmological set of equations now reads
\beq
H^{2}=-\frac{1}{(n-2)}\Big[ 2b_{1}+(n-1)b_{3}+4(n-2) \Big]\Phi^{2}+4 H\Phi +(1-b_{1})P^{2}\delta^n_4+\frac{2 \kappa}{(n-1)(n-2)}\rho \,, \label{Friedmann1}
\eeq 
\beq
\frac{\ddot{a}}{a}=-\frac{\kappa}{(n-1)(n-2)}\Big[ (n-3)\rho+(n-1)p \Big]+\Big[ 2b_{1}+(n-1)b_{3}\Big]\Phi^{2}+2 H \Phi+2 \dot{\Phi} \,, \label{acc2}
\eeq
where $\Phi$ is related to the source field $\phi$ through
\beq
\kappa \phi =-\Big[ 2b_{1}+(n-1)b_{3} \Big]\Phi \,, \quad \chi+\psi=-\phi \,, \label{phiPhi}
\eeq
and we also have that \eqref{chi} and \eqref{psi} hold true, together with $2b_1+(n-1)b_3 \neq0$. In addition, the sources are subject to the conservation laws
\beq
\dot{\rho}_{c}+(n-1)H(\rho_{c}+p_{c})=\frac{1}{2}(\psi-\chi)R_{\mu\nu}u^{\mu}u^{\nu} \,, \label{cont}	
\eeq
\beq
(n-1)(\psi +\chi)(H-Y)=2 (\rho_{c}-\rho) \,, \label{omega}
\eeq
\beq
\dot{\phi}+(n-1)H \phi=\frac{2}{(n-1)}\Big[ -(\rho_{c}-\rho)+(n-1)(p_{c}-p) \Big] \,. \label{phi}
\eeq
Let us now consider $\zeta=0$ and disregard the pseudo-scalar mode, setting $P=0$ (which is consistent with \eqref{Pzeta}, as we can see from our previous analysis). This does not modify the qualitative analysis and the general results we are going to provide. Indeed, by allowing $P\neq0$ one would have to introduce a barotropic equation connecting $P$ to $\Phi$ of the form $P \propto \Phi$ and the presence of $P$ would just introduce a shift in coefficients. 

In order to study the physical cosmology of our Universe, we shall fix $n=4$. With the assumption that the perfect fluid variables of the hyperfluid are related through barotropic equations of state of the usual form,
\beq
p_{c}=w_{c}\rho_{c} \,,
\eeq
\beq
p= w \rho \,,
\eeq
\beq
\rho_{c}=\tilde{w}\rho \,,	
\eeq
where $w_c$, $w$, and $\tilde{w}$ are the associated barotropic indices,	we will now obtain general exact and analytic solutions of the above system. To start with, we first observe the emergence of a perfect square in \eqref{Friedmann1}, which, defining
\beq\label{defxi}
\xi:=H-2 \Phi
\eeq
and
\beq 
b_{0}:=2b_{1}+3b_{3} \,,
\eeq
simplifies to
\beq
\xi^{2}=-\frac{1}{2}b_{0}\Phi^{2}+\frac{\kappa}{3}\rho \,. \label{xi}
\eeq
Furthermore, recalling that $Y=2\Phi$ and using \eqref{phiPhi}, eq. \eqref{omega} can be expressed as
\beq\label{Phixi}
\Phi \xi=\frac{2(\tilde{w}-1)}{3 b_{0}}\kappa \rho \,.
\eeq
Combining the above we obtain
\beq
(\Phi^{2})^{2}-\frac{2 \kappa}{3 b_{0}}\rho \Phi^{2}+\frac{2}{b_{0}}\left[ \frac{2(\tilde{w}-1)\kappa}{3 b_{0}} \right]^{2}\rho^{2}=0 \,. \label{quad}
\eeq
The latter can be seen as a quadratic equation either in $\Phi^{2}$ or in $\rho$. We may see it as a quadratic equation for $\Phi^{2}$. In order to have real solutions, the condition
\beq
1-\frac{8(\tilde{w}-1)^{2}}{b_{0}}\ge 0	
\eeq
have to be satisfied.
Then, it follows that
\beq
\Phi^{2}=\frac{\kappa \rho}{3 b_{0}}\left( 1 \pm \sqrt{1-\frac{8(\tilde{w}-1)^{2}}{b_{0}}} \right) \,.
\eeq
Incidentally, there is yet another constraint that gives us a unique solution. Indeed, from \eqref{xi} we can see that there exists some time period in which the $\Phi^{2}$ component will be dominant over the density $\rho$, and in that region one has
\beq
\xi^{2}\approx -\frac{1}{2}b_{0}\Phi^{2} \,,
\eeq
which demands $b_{0}<0$, otherwise there would be a contradiction. With this result, we extract from \eqref{quad} the unique solution
\beq
\Phi^{2}=\frac{\kappa \rho}{3 b_{0}}\left( 1 - \sqrt{1-\frac{8(\tilde{w}-1)^{2}}{b_{0}}} \right) \,,
\eeq
since the one with the plus sign would certainly give negative $\Phi^{2}$, which is clearly impossible. Setting
\beq
\lambda_{0} :=  1 - \sqrt{1-\frac{8(\tilde{w}-1)^{2}}{b_{0}}}	<0
\eeq 
we have
\beq
\rho=\frac{3 b_{0}}{\kappa \lambda_{0}}\Phi^{2} \,, \label{rhophi}
\eeq
which is positive as expected.
Continuing, we may substitute the latter equation back into \eqref{Phixi} to arrive at (recalling also the definition of $\xi$ given by \eqref{defxi})
\beq
H=\lambda_{1}\Phi \,, \label{HPhi}
\eeq
where we have defined
\beq
\lambda_{1} := 2\left( 1+\frac{\tilde{w}-1}{\lambda_0} \right) \,.
\eeq
Then, plugging \eqref{HPhi} along with \eqref{phiPhi} and \eqref{rhophi} into the evolution equation \eqref{phi}, we find
\beq\label{dotPhi}
\dot{\Phi}=-\lambda_{2}\Phi^{2} \,,
\eeq
where
\beq
\lambda_{2} := \frac{2}{\lambda_{0}}\Big[ 3 \lambda_{0} -2 -3w +\tilde{w}(3 w_{c}+2) \Big] \,. \label{l2}
\eeq
Eq. \eqref{dotPhi} can be trivially integrated to give
\beq
\Phi(t)=\frac{1}{\lambda_{2}t+C_{1}}	\,,
\eeq
where $C_{1}$ is some integration constant to be determined by the initial conditions. With this at hand, we conclude that
\beq
\rho(t)=\frac{3 b_{0}}{\kappa \lambda_{0}}\frac{1}{(\lambda_{2}t+C_{1})^{2}} \,.
\eeq
Furthermore, also \eqref{HPhi} can be now integrated, yielding 
\beq
a(t)=C_{2}(\lambda_{2}t+C_{1})^{\frac{\lambda_{1}}{\lambda_{2}}} \,, \label{a1}
\eeq
where $C_{2}$ is another integration constant which can be fixed from the initial data. From the last equation we see that we have interesting power law solutions for the scale factor which depend on the parameters of the theory.
Note that there is also the second Friedmann equation \eqref{acc2} which must be taken into account. Nevertheless, since it is a byproduct of the first Friedmann equation and the conservation laws, it is not and independent equation and its contribution is already accounted for in the analysis above. Lastly, in order to have a self-consistent theory, all equations must be satisfied and the only one we have not used so far is \eqref{cont}. Substituting all the above results into \eqref{cont} we get the consistency relation 
\beq
4 \tilde{w}\Big[  - 2\lambda_{2}+3(1+w_{c})\lambda_{1} \Big]=(8-b_{0})\Big( 1+3 w + 2\lambda_{0} \Big) \label{consrel}
\eeq 
among the parameters of the theory. Remarkably, the solutions we have provided analytically are exact ones. Some comments regarding our power law solution for the scale factor are now in order. Firstly, note that the evolution of the latter can be either slower or more rapid with respect to the one obtained for conventional forms of matter (such as dust, radiation, etc.{; for a nice review of the various forms of matter in cosmology we refer the reader to \cite{Tsagas:2007yx}}), depending on both the parameters of the theory and the barotropic indices. Given some data one would be able to restrict the possible values of these parameters. Secondly, restrictions on the parameters can be obtained directly from \eqref{a1}. Indeed, given that $a(t)>0$, we must have $C_{2}>0$. Furthermore, in order to have unique real solutions for any value of the ratio $\lambda_{1} / \lambda_{2}$, it must hold that $\lambda_{2}t+C_{1}>0$ for any $t$. For $t=0$ we conclude that $C_{1}>0$, while for late times $\lambda_{2}t$ becomes dominant over $C_{1}$ and the positivity is guaranteed as long as $\lambda_{2}>0$.
With these at hand, given that for some fixed time $t=t_{0}$ the scale factor and the density acquire values $a(t_{0})=a_{0}$ and $\rho(t_{0})=\rho_{0}$, the integration constants are found to be
\beq
C_{1} =\sqrt{\frac{3 b_{0}}{\kappa \lambda_{0}\rho_{0}}} - \lambda_2 t_0 \,, \quad C_{2}=a_{0}\left(\frac{\kappa \lambda_{0}\rho_{0}}{3 b_{0}} \right)^{\frac{\lambda_{1}}{2\lambda_{2}}}
\eeq
and consequently the scale factor can be expressed as
\beq
a(t)=a_{0}\Big[ \tilde{\lambda}(t-t_{0})+1  \Big]^{\frac{\lambda_{1}}{\lambda_{2}}} \,, \label{a}
\eeq
where
\beq
\tilde{\lambda}:=\lambda_{2}\sqrt{\frac{\kappa \lambda_{0}\rho_{0}}{3 b_{0}}}>0 \,. \label{tildelambda}
\eeq
Now, at first sight it seems that there exist only power law solutions for the torsionful system. However, the expansion depends on parameters that could potentially lead to a more rapid expansion. Indeed, given the form of \eqref{l2}, there exists a given configuration among the barotropic indices for which $\lambda_{2}\rightarrow {0^+}$. In this limit the scale factor goes like\footnote{This is easily seen by recalling the identity $\lim_{\mu \to \infty}\Big( 1+\frac{t}{\mu} \Big)^{\mu} = e^{t}$, where in the case at hand we have $\mu=\frac{1}{\lambda_{2}}$.}
\beq
a(t) \propto e^{\lambda_{1}t} \,,
\eeq
which signals an exponential expansion. Although this would of course require fine tuning, it is evident that it represents a possibility. Constraints on both the torsion scalars parameters and the barotropic indices could be obtained by fitting the derived results to some given data. This could also allow one to rule out specific cases and find the allowed equations of state among the hyperfluid variables. Let us discuss here some specific characteristic cases emerging by considering different values of the ratio $\lambda_{1} / \lambda_{2}$.

\paragraph{\textbf{Case $\lambda_1 = 0$, $\lambda_2 \neq 0$:}}

As is can be seen from \eqref{a}, in this case we have a static Universe, $a=a_{0}=\text{const}$. Note that in such a case both the Hubble parameter and its first derivative vanish (i.e. $H=0=\dot{H}$) in agreement with the static nature of the model. {Moreover, besides fulfilling the consistency relation \eqref{consrel}, here we find that the parameters of our theory satisfy the additional relation $\lambda_0=1-\tilde{w}$.}

\paragraph{\textbf{Case $\lambda_{1}\neq 0$, $\lambda_{2} \rightarrow 0^{+}$:}}
As we have already pointed out previously, in  this case the scale factor experiences an exponential growth. Interestingly, in this configuration both $\Phi$ and $\rho$ ``freeze out'' and acquire the constant values
\beq
\Phi= \Phi_{0}\,, \quad \rho = \rho_{0}=\frac{3 b_{0}}{\kappa \lambda_{0}}\Phi^{2}_{0} \,.
\eeq
In a sense, the freezing out of the latter two acts as an effective cosmological constant, 
\beq
\Lambda_{\text{eff}} = \frac{12 (\lambda_{0}+\tilde{w}-1)^{2}}{\lambda_{0}^{2}}\Phi_{0}^{2} \,, 
\eeq
which drives the exponential expansion. In this instance the parameters of the theory satisfy{, besides eq. \eqref{consrel},} $3 \lambda_{0}=2+3 w-\tilde{w}(2+3 w_{c})$.

\paragraph{\textbf{Case $\lambda_{1}/ \lambda_{2}=1 $:}} In this limit eq. \eqref{a} yields
\beq
a(t)=a_{0}\Big[ \tilde{\lambda}(t-t_{0})+1  \Big] \,.
\eeq
In this case we notice a Milne-like expansion and in particular when $a_{0}t_{0}=1=a_{0}\tilde{\lambda}$  the behaviour is identical with that of a Milne Universe {\cite{milne}} (i.e. $a=t$).
{The parameters also fulfill $2 \lambda_0 = 3 w + 1 - \tilde{w}(1+3 w_c)$.}

\paragraph{\textbf{Case $\lambda_{1}/ \lambda_{2}=1/2 $:}} For such a configuration we have
\beq
a(t)=a_{0}\Big[ \tilde{\lambda}(t-t_{0})+1  \Big]^{\frac{1}{2}} \,,
\eeq
{indicating} a radiation-like expansion. The parameters are related through {$\lambda_0 = 3 ( w - \tilde{w} w_c )$.}

\paragraph{\textbf{Case $\lambda_{1}/ \lambda_{2}=2/3 $:}} In this  case we get
\beq
a(t)=a_{0}\Big[ \tilde{\lambda}(t-t_{0})+1  \Big]^{\frac{2}{3}} \,,
\eeq
from which we conclude that the net effect is similar to that of dust in comparison to the solutions predicted by the Standard Cosmological Model. Here, the model parameters satisfy {$3 \lambda_0 = 1 + 6 w - \tilde{w} (1+6 w_c)$.}

\paragraph{\textbf{Case $\lambda_{1}/ \lambda_{2}=1/3 $:}} In this case we have
\beq
a(t)=a_{0}\Big[ \tilde{\lambda}(t-t_{0})+1  \Big]^{\frac{1}{3}} \,.
\eeq
The above now indicates a correspondence with a stiff matter dominated Universe and the parameters obey {$3 w=1 - \tilde{w}(1+3w_c)$.}

The above represent only some very specific correspondences with respect to standard cosmology. In particular, we should note that, depending on the parameter space, our solutions also allow for an accelerated growth when $\lambda_{1}/\lambda_{2}>0$, in contrast to the standard picture where conventional matter always causes a slow expansion. We see therefore that torsion changes this picture dramatically and allows for interesting possibilities.

\paragraph*{\textbf{The case $\tilde{w}=1$:}} 

Note that in all of the above considerations we have assumed that $\tilde{w}\neq 1$. In the special case for which $\tilde{w}=1$, the total (canonical) density does not receive contributions from the hypermomentum part, i.e. $\rho_{c}=\rho$. Let us analyze this case further. From {\eqref{Phixi}, recalling also the definition for $\xi$ given in \eqref{defxi},} for $\tilde{w}=1$ we get that either $\Phi=0$ or $H=2 \Phi$. The former {represents a trivial solution}, so we shall consider the latter possibility. Then, for $H=2 \Phi$ eq. \eqref{quad} becomes
\beq
\rho=\frac{3 b_{0}}{2 \kappa}\Phi^{2} \,.
\eeq
Interestingly, in this case the physical restriction $\rho>0$ demands that $b_{0}>0$. Substituting the above results into the conservation law {\eqref{phi} and using also \eqref{phiPhi}} it follows that
\beq
\dot{\Phi}=-3\Big[ 2+(w_{c}-w)\Big] \Phi^{2} \,,
\eeq
or, equivalently,
\beq
\dot{H}=-\frac{w_{0}}{2}H^{2} \,, \label{decelparam}
\eeq
where $w_{0}:=6+3(w_{c}-w)$. {Again, the latter can be} trivially integrated to give
\beq
a(t)=C_{2}\Big( \frac{w_{0}}{2}t+C_{1}\Big)^{-\frac{2}{w_{0}}} \,.
\eeq
Then, {classifications similar to the ones obtained in the case $\tilde{w}\neq 1$ follow.} However, let us observe that the solution for the scale factor here is independent of the parameter $b_{0}$ and{, as a consequence, it does not depend on} the coefficients of the additional quadratic torsion terms. Furthermore, {in the case of} a hypermomentum preserving hyperfluid \cite{Iosifidis:2020gth}, {for which $\rho_{c}=\rho$ and $p_{c}=p$,} we get the unique solution for the scale factor
\beq
{a(t)= C_2 \Big( 3 t+ C_1 \Big)^{-\frac{1}{3}} \,,}
\eeq
{which is perfectly allowable since in our theory the torsion contributions modify the early time cosmology.}

\section{Conclusions}\label{conclusions}

We have considered a quadratic torsionful gravity theory in $n$ spacetime dimensions in the presence of a perfect hyperfluid and we have developed and studied its cosmology. The gravitational action we considered is an extension of the Einstein-Cartan theory including also the three allowed torsion parity even squared terms. The inclusion of the quadratic terms turns out to be most important as it solves the problem of indeterminacy\footnote{This indeterminacy is related to the fact that the Palatini tensor is traceless when contracted in its first and second index. In an FLRW space this trivializes the hypermomentum conservation law. The presence of the quadratic terms resolves exactly this problem.} that one faces when only the Ricci scalar is included into the gravitational action. As for the matter part we considered the presence of a perfect hyperfluid which has been recently developed. The metric and the connection have been considered as independent variables and the equations of motion of the theory have been derived in this setup. We have studied the cosmology of the theory considering the usual FLRW background and discussed the non-Riemannian torsion driven degrees of freedom within the latter. We have then analyzed in detail the conservation laws of the perfect hyperfluid and also the torsion modified Friedman equations for our theory. Remarkably, for this seemingly complicated model, we have been able to provide exact analytic cosmological solutions, finding in particular power law solutions for the scale factor which depend on the parameters of the theory. Under certain circumstances the expansion can be very rapid, i.e. exponential-like. Under a general perspective, our solutions for the scale factor provide generalizations of the usual dust, radiation, or general barotropic perfect fluid solutions of the standard cosmology. The reason for this generalized possibility lies in the inclusion of the hypermomentum degrees of freedom, which, having a direct association with the intrinsic characteristics of the material fluid, modify the net expansion. In this sense the microstructure of the fluid alters the expansion rate in a non-trivial way and provides new interesting cosmological results.

{We have also discussed some specific characteristic cases emerging by considering different values of the ratio $\lambda_{1} / \lambda_{2}$. In particular, for $\lambda_1/\lambda_2=0$ we have found a static Universe, while for $\lambda_1/\lambda_2=1$ we have obtained a Milne-like expansion. In the case $\lambda_2 \rightarrow 0^+$, corresponding to an exponential growth, we have also derived the effective cosmological constant, whereas the cases $\lambda_1/\lambda_2 = 1/2$, $\lambda_1/\lambda_2 = 2/3$, and $\lambda_1/\lambda_2 = 1/3$ correspond, respectively, to a radiation-like expansion, dust effects, and stiff matter dominated Universe. For each case we have also found further bound on the parameters. Interestingly, depending on the parameter space, our torsionful solutions allow for an accelerated growth when $\lambda_{1}/\lambda_{2}>0$, in contrast to the standard picture where conventional matter always causes a slow expansion. Finally, we have analyzed the special case $\tilde{w}=1$, namely the one in which the hypermomentum sector does not contribute to the total (canonical) density ($\rho_c=\rho$). We have derived the expression for the scale factor also in this setup, observing that for the particular case of a hypermomentum preserving hyperfluid ($\rho_c=\rho$, $p_c=p$) the solution is in fact unique.}

In closing, let us note that there exist many possible ways to extend our present study. For instance one could also add the quadratic curvature terms and investigate the cosmology of the Poincaré theory in the presence of the perfect hyperfluid. It would also be interesting to see which would be the effect of additional parity odd quadratic torsion terms in a more generalized setting.
Finally, a probably more ambitious work would be to generalize the setup of the present study by allowing for a non-vanishing non-metricity and subsequently obtain the full cosmology of the resulting quadratic MAG theory. Some work is currently in progress on this point. 

\vspace{0.5cm}

\begin{center}
\textbf{Acknowledgements}
\end{center}

{We would like to thank very much Christos  Tsagas for some helpful discussions and useful comments regarding the interpretation of the solutions.}
D.I. acknowledges: This research is co-financed by Greece and the European Union (European Social Fund - ESF) through the Operational Programme `Human Resources Development, Education and Lifelong Learning' in the context of the project ``Reinforcement of Postdoctoral Researchers - 2nd Cycle'' (MIS-5033021), implemented by the State Scholarships Foundation (IKY).
L.R. would like to thank the Department of Applied Science and Technology of the Polytechnic University of Turin, and in particular Laura Andrianopoli and Francesco Raffa, for financial support.

\appendix	
	
\section{Useful formulas}\label{appa}

Our convention for the metric signature is mostly plus. In particular, the $n=4$ metric signature is $(-,+,+,+)$.
Some useful formulas that we have derived and exploited in our calculations are the following:
\begin{align}
& S_{\mu\nu\alpha}u^{\alpha}=0 \,, \\
& u^{\mu}S_{\mu\alpha\beta}=-\Phi h_{\alpha\beta} \,, \\
& {S_{\nu}}^{\alpha\beta}u_{\alpha}h_{\mu\beta}=\Phi h_{\mu\nu} \,, \\
& S_{\mu\alpha\beta}h^{\alpha\beta}=S_{\mu}=(n-1)\Phi u_{\mu} \,, \\
& S_{\mu\alpha\beta}{S_{\nu}}^{\alpha\beta}=\Phi^{2}\Big[ (n-1)u_{\mu}u_{\nu}-h_{\mu\nu} \Big] +2 P^{2} \delta^n_4 h_{\mu\nu} \,, \\
& S_{\mu\alpha\beta}{S_{\nu}}^{\beta\alpha}=(n-1)\Phi^{2}u_{\mu}u_{\nu}-2 P^{2} \delta^n_4 h_{\mu\nu} \,, \\
& S_{\mu}S_{\nu}=(n-1)^{2}\Phi^{2}u_{\mu}u_{\nu} \,, \\
& S_{\alpha\beta\mu}{S^{\alpha\beta}}_{\nu}=(-2 \Phi^{2}+2 P^{2} \delta^n_4 )h_{\mu\nu} \,, 
\end{align}
\begin{align}
& B_{\mu\nu}= b_{1}(2S_{\nu\alpha\beta}{S_{\mu}}^{\alpha\beta}-S_{\alpha\beta\mu} {S^{\alpha\beta}}_{\nu})-b_{2}S_{\nu\alpha\beta}{S_{\mu}}^{\beta\alpha}+b_{3}S_{\mu}S_{\nu} \nonumber \\
& \phantom{B_{\mu\nu}} = (n-1)\Phi^{2}u_{\mu}u_{\nu}\Big[ 2 b_{1}-b_{2}+(n-1)b_{3}\Big] +2 (b_{1}+b_{2})P^{2} \delta^n_4 h_{\mu\nu} \,, \\
& B_{\mu\nu}u^{\mu}u^{\nu}=(n-1)\Phi^{2}	\Big[ 2 b_{1}-b_{2}+(n-1)b_{3}\Big] \,, \\
& B_{\mu\nu}h^{\mu\nu}= 2(b_{1}+b_{2})(n-1)P^{2} \delta^n_4 \,, \\
& B:=B_{\mu\nu}g^{\mu\nu}=(n-1) \Big\lbrace 2 (b_{1}+b_{2})P^{2} \delta^n_4-\Big[ 2 b_{1}-b_{2}+(n-1)b_{3}\Big]\Phi^{2} \Big\rbrace \,, \\
& B_{ij} = 2(b_{1}+b_{2})P^{2} \delta^n_4 g_{ij} \,, \\
& B_{00}=(n-1)\Phi^{2}\Big[ 2 b_{1}-b_{2}+(n-1)b_{3}\Big] \,.
\end{align}
Furthermore, regarding the metric field equations, taking the trace of \eqref{metricf} and plugging back the result into the latter we get
\beq
R_{\mu\nu}u^{\mu}u^{\nu}=\kappa \left[ \frac{1}{(n-2)}T+T_{\mu\nu}u^{\mu}u^{\nu}\right]-(n-1)\Big[ 2b_{1}-b_{2}+(n-1)b_{3}\Big]\Phi^{2} \,.
\eeq	
Then, upon use of \eqref{metrenmomform}, together with the trace of the latter, we are left with	
\beq
R_{\mu\nu}u^{\mu}u^{\nu}=\frac{\kappa}{(n-2)} \Big[ (n-3)\rho +(n-1)p \Big]-(n-1)\Big[ 2b_{1}-b_{2}+(n-1)b_{3}\Big] \Phi^{2} \,,
\eeq	
which has been used to write the Friedmann equations in the main text.

Let us also give the relevant hypermomentum contractions, that are
\begin{align}
& h^{\alpha \mu} \Delta_{\alpha \mu \nu} = (n-1) \phi u_\nu \,, \\
& h^{\alpha \nu} \Delta_{\alpha \mu \nu} = (n-1) \chi u_\mu \,, \\
& h^{\mu \nu} \Delta_{\alpha \mu \nu} = (n-1) \psi u_\alpha \,, \\
& \varepsilon^{\alpha \mu \nu \lambda} \Delta_{\alpha \mu \nu} = -6 u^\lambda \zeta \delta^n_4 \,, \\
& u^\alpha u^\mu u^\nu \Delta_{\alpha \mu \nu} = - \omega \,,
\end{align}	
which can be also inverted, yielding
\begin{align}
& \phi = - \frac{1}{(n-1)} h^{\alpha \mu} u^\nu \Delta_{\alpha \mu \nu} \,, \\
& \chi = - \frac{1}{(n-1)} h^{\alpha \nu} u^\mu \Delta_{\alpha \mu \nu} \,, \\
& \psi = - \frac{1}{(n-1)} h^{\mu \nu} u^\alpha \Delta_{\alpha \mu \nu} \,, \\
& \zeta = \frac{1}{6} \varepsilon^{\alpha \mu \nu \lambda} \Delta_{\alpha \mu \nu} u_\lambda \delta^n_4 \,, \\
& \omega = - u^\alpha u^\mu u^\nu \Delta_{\alpha \mu \nu} \,,
\end{align}
where we recall that $\omega=0$ for vanishing non-metricity, together with the explicit form and contractions of the Palatini tensor in our cosmological setup, which read
\begin{align}
& P_{\alpha \mu \nu} = 4(n-2) \Phi u_{[\alpha} h_{\mu] \nu} - 2 \varepsilon_{\alpha \mu \nu \rho} u^\rho P \delta^n_4 \,, \\
& h^{\alpha \mu} P_{\alpha \mu \nu} = 0 \,, \\
& h^{\alpha \nu} P_{\alpha \mu \nu} = -2 (n-1)(n-2) \Phi u_\mu \,, \\
& h^{\mu \nu} P_{\alpha \mu \nu} = 2 (n-1)(n-2) \Phi u_\alpha \,, \\
& \varepsilon^{\alpha \mu \nu \lambda} P_{\alpha \mu \nu} = 12 P u^\lambda \delta^n_4 \,, \\
& u^\alpha u^\mu u^\nu P_{\alpha \mu \nu} = 0 \,.
\end{align}
Clearly, if non-vanishing non-metricity were also allowed, the above equations would be modified.

%%%%%%%%%%%%%%%%%%%%%%%%%%%%	


\begin{thebibliography}{99}

\bibitem{Clifton:2011jh}
T.~Clifton, P.~G.~Ferreira, A.~Padilla and C.~Skordis,
``Modified Gravity and Cosmology,''
Phys. Rept. \textbf{513} (2012), 1-189
[arXiv:1106.2476 [astro-ph.CO]].

\bibitem{eisenhart} 
L.~P.~Eisenhart,
``Non-Riemannian geometry,''
American Mathematical Society - Colloquium Publications, 1927, Volume VIII, 184 pages.

\bibitem{schouten}
J.~A.~Schouten,
``Ricci-calculus: an introduction to tensor analysis and its geometrical applications,''
Springer Science $\&$ Business Media, 2013, Volume 10.

\bibitem{Puetzfeld:2004yg}
D.~Puetzfeld,
``Status of non-Riemannian cosmology,''
New Astron. Rev. \textbf{49} (2005), 59-64
[arXiv:gr-qc/0404119 [gr-qc]]

\bibitem{Puetzfeld:2005af}
D.~Puetzfeld,
``Prospects of non-Riemannian cosmology,''
eConf \textbf{C041213} (2004), 1221
[arXiv:astro-ph/0501231 [astro-ph]].

\bibitem{Klemm:2018bil}
D.~S.~Klemm and L.~Ravera,
``Einstein manifolds with torsion and nonmetricity,''
Phys. Rev. D \textbf{101} (2020) no.4, 044011
[arXiv:1811.11458 [gr-qc]].

\bibitem{Klemm:2019izb}
D.~S.~Klemm and L.~Ravera,
``Supersymmetric near-horizon geometry and Einstein-Cartan-Weyl spaces,''
Phys. Lett. B \textbf{793} (2019), 265-270
[arXiv:1904.03681 [hep-th]].

\bibitem{Klemm:2020mfp}
S.~Klemm and L.~Ravera,
``An action principle for the Einstein\textendash{}Weyl equations,''
J. Geom. Phys. \textbf{158} (2020), 103958
[arXiv:2006.15890 [hep-th]].

\bibitem{Klemm:2020gfm}
S.~Klemm and L.~Ravera,
``Schr\"odinger connection with selfdual nonmetricity vector in 2+1 dimensions,''
[arXiv:2008.12740 [hep-th]].

\bibitem{Iosifidis:2020dck}
D.~Iosifidis and L.~Ravera,
``Parity Violating Metric-Affine Gravity Theories,''
[arXiv:2009.03328 [gr-qc]].

\bibitem{Kranas:2018jdc}
D.~Kranas, C.~G.~Tsagas, J.~D.~Barrow and D.~Iosifidis,
``Friedmann-like universes with torsion,''
Eur. Phys. J. C \textbf{79} (2019) no.4, 341
[arXiv:1809.10064 [gr-qc]].

\bibitem{Pereira:2019yhu}
S.~H.~Pereira, R.~d.~C.~Lima, J.~F.~Jesus and R.~F.~L.~Holanda,
``Acceleration in Friedmann cosmology with torsion,''
Eur. Phys. J. C \textbf{79} (2019) no.11, 950
[arXiv:1906.07624 [gr-qc]].

\bibitem{Barrow:2019bvx}
J.~D.~Barrow, C.~G.~Tsagas and G.~Fanaras,
``Friedmann-like universes with weak torsion: a dynamical system approach,''
Eur. Phys. J. C \textbf{79} (2019) no.9, 764
[arXiv:1907.07586 [gr-qc]].

\bibitem{Guimaraes:2020drj}
T.~M.~Guimar\~aes, R.~d.~C.~Lima and S.~H.~Pereira,
``Cosmological inflation driven by a scalar torsion function,''
[arXiv:2011.13906 [gr-qc]].

\bibitem{Iosifidis:2020zzp}
D.~Iosifidis,
``Cosmic Acceleration with Torsion and Non-metricity in Friedmann-like Universes,''
Class. Quant. Grav. \textbf{38} (2021) no.1, 015015
[arXiv:2007.12537 [gr-qc]].

\bibitem{aldrovandi}
R.~Aldrovandi and J.~G.~Pereira,
``Teleparallel gravity: an introduction,''
Springer Science $\&$ Business
Media, 2012, Volume 173.

\bibitem{Nester:1998mp}
J.~M.~Nester and H.~J.~Yo,
``Symmetric teleparallel general relativity,''
Chin. J. Phys. \textbf{37} (1999), 113
[arXiv:gr-qc/9809049 [gr-qc]].

\bibitem{BeltranJimenez:2018vdo}
J.~Beltr\'an Jim\'enez, L.~Heisenberg and T.~S.~Koivisto,
``Teleparallel Palatini theories,''
JCAP \textbf{08} (2018), 039
[arXiv:1803.10185 [gr-qc]].

\bibitem{Jimenez:2019ghw}
J.~Beltr\'an Jim\'enez, L.~Heisenberg, D.~Iosifidis, A.~Jim\'enez-Cano and T.~S.~Koivisto,
``General teleparallel quadratic gravity,''
Phys. Lett. B \textbf{805} (2020), 135422
[arXiv:1909.09045 [gr-qc]].

\bibitem{Hehl:1994ue}
F.~W.~Hehl, J.~D.~McCrea, E.~W.~Mielke and Y.~Ne'eman,
``Metric affine gauge theory of gravity: Field equations, Noether identities, world spinors, and breaking of dilation invariance,''
Phys. Rept. \textbf{258} (1995), 1-171
[arXiv:gr-qc/9402012 [gr-qc]].

\bibitem{Hehl:1999sb}
F.~W.~Hehl and A.~Macias,
``Metric affine gauge theory of gravity. 2. Exact solutions,''
Int. J. Mod. Phys. D \textbf{8} (1999), 399-416
[arXiv:gr-qc/9902076 [gr-qc]].

\bibitem{Iosifidis:2019jgi}
D.~Iosifidis, \\
``Metric-Affine Gravity and Cosmology/Aspects of Torsion and non-Metricity in Gravity Theories,''
[arXiv:1902.09643 [gr-qc]].	

\bibitem{Hehl:1976kt}
F.~W.~Hehl, G.~D.~Kerlick and P.~Von Der Heyde,
``On Hypermomentum in General Relativity. 1. The Notion of Hypermomentum,''
Z. Naturforsch. A \textbf{31} (1976), 111-114

\bibitem{Weyssenhoff:1947iua}
J.~Weyssenhoff and A.~Raabe,
``Relativistic dynamics of spin-fluids and spin-particles,''
Acta Phys. Polon. \textbf{9} (1947), 7-18

\bibitem{Obukhov:1993pt}
Y.~N.~Obukhov and R.~Tresguerres,
``Hyperfluid: A Model of classical matter with hypermomentum,''
Phys. Lett. A \textbf{184} (1993), 17-22
[arXiv:gr-qc/0008013 [gr-qc]].

\bibitem{Obukhov:1996mg}
Y.~N.~Obukhov,
``On a model of an unconstrained hyperfluid,''
Phys. Lett. A \textbf{210} (1996), 163-167
[arXiv:gr-qc/0008014 [gr-qc]].

\bibitem{Babourova:1995fv}
O.~V.~Babourova and B.~N.~Frolov,
``The Variational theory of perfect fluid with intrinsic hypermomentum in space-time with nonmetricity,''
[arXiv:gr-qc/9509013 [gr-qc]].

\bibitem{Babourova:2004xx}
O.~V.~Babourova and B.~N.~Frolov,
``Perfect hypermomentum fluid: Variational theory and equations of motion,''
Int. J. Mod. Phys. A \textbf{13} (1998), 5391-5407
[arXiv:gr-qc/0405124 [gr-qc]].

\bibitem{Ray:1982qr}
J.~R.~Ray and L.~L.~Smalley,
``Spinning Fluids in the Einstein-cartan Theory,''
Phys. Rev. D \textbf{27} (1983), 1383

\bibitem{Smalley}
L.~L.~Smalley,
``Fluids with spin and twist,''
Journal of Mathematical Physics \textbf{36}, 778 (1995)

\bibitem{Iosifidis:2020gth}
D.~Iosifidis, \\
``Cosmological Hyperfluids, Torsion and Non-metricity,''
Eur. Phys. J. C \textbf{80} (2020) no.11, 1042
[arXiv:2003.07384 [gr-qc]]

\bibitem{Iosifidis:2020upr}
D.~Iosifidis,
``Non-Riemannian Cosmology: The role of Shear Hypermomentum,''
[arXiv:2010.00875 [gr-qc]].

\bibitem{Iosifidis:2021nra}
D.~Iosifidis,
``The Perfect Hyperfluid of Metric-Affine Gravity: The Foundation,''
[arXiv:2101.07289 [gr-qc]].

\bibitem{Baekler:2011jt}
P.~Baekler and F.~W.~Hehl,
``Beyond Einstein-Cartan gravity: Quadratic torsion and curvature invariants with even and odd parity including all boundary terms,''
Class. Quant. Grav. \textbf{28} (2011), 215017
[arXiv:1105.3504 [gr-qc]].

\bibitem{Obukhov:2014nja}
Y.~N.~Obukhov and D.~Puetzfeld,
``Conservation laws in gravity: A unified framework,''
Phys. Rev. D \textbf{90} (2014) no.2, 024004
[arXiv:1405.4003 [gr-qc]].

\bibitem{Iosifidis:2018zwo}
D.~Iosifidis and T.~Koivisto,
``Scale transformations in metric-affine geometry,''
[arXiv:1810.12276 [gr-qc]].

\bibitem{mtsamparlis}
Michael Tsamparlis, 
``Cosmological principle and torsion,'' Phys. Lett. A, \textbf{75} (1979) 27–28.	

\bibitem{Tsagas:2007yx}
C.~G.~Tsagas, A.~Challinor and R.~Maartens,
``Relativistic cosmology and large-scale structure,''
Phys. Rept. \textbf{465} (2008), 61-147
[arXiv:0705.4397 [astro-ph]].

\bibitem{milne}
E.~A.~Milne, 
``Relativity, Gravitation and World structure,'' 
Oxford University Press, Oxford, 1935.

\end{thebibliography}
\end{document}